\documentstyle[12pt,epsfig]{article}
\begin{document}
\title{\vspace{-15mm}
       \vspace{2.5cm}
       {\normalsize \hfill
       \begin{tabbing}
       \`\begin{tabular}{l}
         UFR-HEP/00/06\\
         KEK-TH/00/699\\
        June 2000 \\
	\end{tabular}
       \end{tabbing} }
       \vspace{2mm}
Note on tree-level unitarity in the General \\ 
Two Higgs Doublet Model.}
\vspace{2mm}
\author{A.G Akeroyd$^a$,  A. Arhrib$^{ b,c}$
,  E. Naimi$^{b, c}$\\[3mm] 
 {\normalsize \em a: KEK Theory Group, Tsukuba,} \\
{\normalsize \em Ibaraki, Japan 305-0801 } \\[3mm] 
 {\normalsize \em b: D\'epartement de Math\'ematiques, 
 Facult\'e des Sciences et Techniques,} \\
{\normalsize \em  B.P 416 Tanger Morocco } \\[3mm] 
 {\normalsize \em c: UFR--Physique des Hautes Energies, 
 Facult\'e des Sciences}\\
{\normalsize \em P.O Box 1014, Rabat--Morocco}  }
\setcounter{footnote}{4} 
\maketitle 
\begin{abstract}
Tree-level unitarity constraints on the masses of the Higgs bosons
in the general Two Higgs Doublet Model (THDM) are studied. 
We first consider the case where the Higgs potential is invariant 
under a discrete symmetry transformation, and derive strong
constraints on the mass of the lightest CP-even Higgs boson ($M_h$)
as a function of $\tan\beta$. We then show that the inclusion of 
the discrete symmetry breaking term weakens the mass bounds considerably.
It is suggested that a measurement of $M_h$ and $\tan\beta$ may enable 
discrimination between the two Higgs potentials.

\end{abstract}

\newpage
\section{Introduction}
The Minimal Standard Model (MSM) of electroweak
interactions \cite{Wein} is in complete agreement 
with all precision experimental data (LEPII,Tevatron, SLD), 
with the notable exception of neutrino oscillation experiments 
\cite{Kam}.  In this minimal version there is one complex $SU(2)_L\otimes U(1)$ 
doublet which provides mass for the fermions and gauge bosons.
After electroweak symmetry breaking three of the four real degrees of freedom
initially present in the Higgs doublet become the 
longitudinal components of the gauge bosons $W^\pm$, $Z$,
leaving one degree of freedom which manifests itself as a physical 
particle ($\phi^0$) \cite{Higgs}. The Higgs mass ($M_{\phi^0}$) 
is not fixed by the 
model, although constraints on $M_{\phi^0}$ can be obtained by making
additional theoretical assumptions. These include the
unitarity  bound ($M_{\phi^0}< 870$ GeV) \cite{uni} and the
triviality bound \cite{dashen}. So far no experimental information on the nature 
of the Higgs particle has been found, and from the negative searches in the 
Higgsstrahlung channel at LEPII the lower bound $M_{\phi^0} > 107.7 $ GeV 
\cite{LEPII} has been derived.

In recent years there has been growing interest in the study of extended 
Higgs sectors with more than one doublet \cite{Gun}. 
The simplest extension of the MSM is the Two Higgs 
Doublet Model (THDM), which is formed by adding an extra complex
$SU(2)_L\otimes U(1)_Y$ scalar doublet to the MSM Lagrangian. 
Motivations for such a structure include CP--violation in the Higgs 
sector, supersymmetry, and a possible solution to the cosmological 
domain wall problem \cite{preskill}. In particular, the Minimal 
Supersymmetric Standard Model (MSSM) 
\cite{Gun} takes the form of a constrained THDM.

The most general THDM scalar potential which is renormalizable,
gauge invariant and CP invariant depends on ten parameters,
but such a potential can still break CP spontaneously \cite{Lee}. 
In order to ensure that tree-level flavour changing neutral currents
are eliminated, a discrete symmetry ($\Phi_i\to -\Phi_i$, where 
$\Phi_i$ is a scalar doublet) may be imposed on the lagrangian
\cite{sym}, which reduces the number of free parameters to 6.
The resulting potential was considered in \cite{sher}, and is 
referred to as $V_A$ in \cite{brucher}. 
We shall be concerned with the potential described in \cite{Gun} 
which is equivalent to $V_A$ plus a term which breaks the discrete symmetry
(parametrized by $\lambda_5$) and contains 7 free parameters. 
Such a potential does not break CP spontaneously or explicitly 
\cite{sher},\cite{branco} provided that all the parameters are real. 

We note here that tree-level unitarity constraints for 
the THDM scalar potential $V_A$ were studied in 
\cite{malampi, kanemura}. When deriving constraints from unitarity, 
\cite{malampi} considered only seven elastic scattering processes 
$S_1S_2\to S_1 S_2$ (where $S_i$ is a Higgs scalar)
while \cite{kanemura} considered 
a larger scattering ($S$) matrix. In \cite{kanemura}, upper bounds 
on the Higgs masses were derived, in particular $M_h\le 410$ GeV 
for $\tan\beta=1$, with the bound becoming stronger as $\tan\beta$ increases.
We improve those studies for $V_A$ by including the full scalar $S$
matrix which includes channels which were absent in \cite{kanemura},
and we also show graphically the strong correlation between 
$M_h$ and $\tan\beta$.

To our knowledge such unitarity constraints
have not been considered for the case of $\lambda_5\ne 0$ and 
this is the principal aim of this note. We shall see that the presence of
a non-zero $\lambda_5$ can significantly weaken the unitarity bounds found in 
\cite{malampi, kanemura}. 

The paper is organized as follows. In Section 2 we give a short review of 
the THDM potential and explain the unitarity approach we 
will be using. In Section 3 we present our numerical results for the
cases of $\lambda_5=0$ and $\ne 0$, while Section 4 contains our conclusions.

\renewcommand{\theequation}{2.\arabic{equation}}
\setcounter{equation}{0}
\section*{2. Scalar potential and unitarity constraint}
It has been shown 
\cite{Geo} that the most general THDM scalar potential which is 
invariant under $SU(2)_L\otimes U(1)_Y$ and conserves CP is given by:
\begin{eqnarray}
 V(\Phi_{1}, \Phi_{2})& & =  \lambda_{1} ( |\Phi_{1}|^2-v_{1}^2)^2
+\lambda_{2} (|\Phi_{2}|^2-v_{2}^2)^2+
\lambda_{3}((|\Phi_{1}|^2-v_{1}^2)+(|\Phi_{2}|^2-v_{2}^2))^2 
+\nonumber\\ [0.2cm]
&  & \lambda_{4}(|\Phi_{1}|^2 |\Phi_{2}|^2 - |\Phi_{1}^+\Phi_{2}|^2  )+
\lambda_{5} (Re(\Phi^+_{1}\Phi_{2})
-v_{1}v_{2})^2+ \lambda_{6} [Im(\Phi^+_{1}\Phi_{2})]^2 
\label{higgspot}
\end{eqnarray}
where $\Phi_1$ and $\Phi_2$ have weak hypercharge Y=1, $v_1$ and
$v_2$ are respectively the vacuum
expectation values of $\Phi_1$ and $\Phi_2$ and the $\lambda_i$'s
are real--valued parameters. 

$$\Phi_i=\left( \begin{array}{c}
\varphi_i^+\\
v_i+\frac{h_i + iz_i}{\sqrt{2}}
\end{array}\right) $$
This potential violates the discrete symmetry
$\Phi_i\to -\Phi_i$ softly by the dimension 2 term
$\lambda_5 Re(\Phi^+_{1}\Phi_{2})$ and has the same 
general structure of the scalar potential of the MSSM.
One can prove easily that for $\lambda_5=0$  
the exact symmetry $\Phi_i \to -\Phi_i$ is recovered.

After electroweak symmetry breaking, the W and Z gauge 
bosons acquire masses  
given by  $m_W^2=\frac{1}{2}g^2 v^2$ and 
$m_Z^2= \frac{1}{2}(g^2 +g'^2) v^2$,
where $g$ and $g'$ are the $SU(2)_{weak}$ and 
$U(1)_Y$ gauge couplings and
$ v^2= v_1^2 + v_2^2$. The combination $v_1^2 + v_2^2$ 
is thus fixed by the electroweak 
scale through $v_1^2 + v_2^2=(2\sqrt{2} G_F)^{-1}$, 
and we are left with 7 free parameters in eq.(\ref{higgspot}), 
namely the $(\lambda_i)_{i=1,\ldots,6}$'s and 
$\tan\beta=v_2/v_1$. Meanwhile,  three of the eight degrees 
of freedom  of the two Higgs doublets correspond to 
the 3 Goldstone bosons ($G^\pm$, $G^0$) and  
the remaining five become physical Higgs bosons: 
$H^0$, $h^0$ (CP--even), $A^0$ (CP--odd)
and $H^\pm$. Their masses are obtained as usual 
by the shift $\Phi_i\to \Phi_i + v_i$ \cite{Gun}. 
After generating the scalar masses in term of scalar parameters
$\lambda_i$, using  straightforward algebra one can express all the 
$\lambda_i$ as functions of the physical masses:
\begin{eqnarray}
& & \lambda_4=\frac{g^2}{2 m^2_W} m_{H^\pm}^2 \ \ ,  \ \
\lambda_6=\frac{g^2}{2 m^2_W} m_{A}^2  \label{lambda46} 
\ \ , \ \  \lambda_3=\frac{g^2}{8 m^2_W}
\frac{\mbox{s}_\alpha \mbox{c}_\alpha}{ \mbox{s}_\beta \mbox{c}_\beta } 
(m_H^2-m_h^2)\ -
\  \frac{\lambda_5}{4} \label{lambda5}  \\
& & \lambda_1 = \frac{g^2}{8 \mbox{c}_\beta^2 m^2_W} 
[ \mbox{c}_\alpha^2 m^2_H+
\mbox{s}_\alpha^2 m^2_h -
\frac{\mbox{s}_\alpha \mbox{c}_\alpha}{\tan\beta}(m^2_H - m^2_h)] 
 -\frac{\lambda_5}{4}(-1 + \tan^2\beta) \label{lambda1} \\ & &
\lambda_2 = \frac{g^2}{8 \mbox{s}_\beta^2 m^2_W} [ \mbox{s}_\alpha^2 m^2_H+
\mbox{c}_\alpha^2 m^2_h -
\mbox{s}_\alpha \mbox{c}_\alpha \tan\beta(m^2_H - m^2_h)] 
 -\frac{\lambda_5}{4}(-1 + \frac{1}{\tan^2\beta} ) \label{lambda1} 
\end{eqnarray}
The angle $\beta$ diagonalizes both the CP--odd and charged scalar
mass matrices, leading to the physical states $H^\pm$ and $A^0$.
The angle $\alpha$ diagonalizes the  CP--even mass matrix
leading to the physical states $H^0$, $h^0$.

We are free to take as 7 independent parameters 
$(\lambda_i)_{i=1,\ldots , 6}$ and $\tan\beta$
or equivalently the four scalar masses, $\tan\beta$, $\alpha$
and one of the $\lambda_i$. In what 
follows we will take $\lambda_5$ as a free parameter.

To constrain the scalar potential parameters one can demand 
that tree-level unitarity is preserved in a variety of scattering
processes. This corresponds to the requirement 
that the $J=0$ partial waves ($a_0$) for 
scalar-scalar and gauge boson scalar scattering
satisfy $ | a_0 | < 1/2$ in the high-energy limit. 
At very high-energy, 
the equivalence theorem \cite{equivalence} states that
the amplitude of a scattering process involving longitudinal gauge 
bosons $V_{\mu}^{\pm, 0}$
may be approximated by the scalar amplitude in which gauge bosons are 
replaced by their corresponding Goldstone bosons $G^{\pm, 0}$. 
We conclude that unitarity constraints can be implemented
by solely considering pure scalar scattering.

In very high energy collisions, it can be shown that 
the dominant contribution to the amplitude of the two-body scattering 
$S_1 S_2 \to S_3 S_4$ is the one which is mediated by the quartic 
coupling. Those contributions mediated by trilinear couplings
are suppressed on dimensional grounds. 
Therefore the unitarity constraint $|a_0|\leq 1/2$ reduces to 
the following constraint on the quartic coupling,
$|Q( S_1 S_2  S_3 S_4)|\leq 8 \pi$.
In what follows our attention will be focussed on the  
the quartic couplings.

In order to derive the unitarity constraints on the scalar masses
we will adopt the technique introduced in \cite{kanemura}.
It has been shown in previous works \cite{our} 
that the quartic scalar vertices written
in terms of physical fields $H^\pm$, $G^\pm$, $h^0$, $H^0$, $A^0$ 
and $G^0$, are very complicated functions of $\lambda_i$, 
$\alpha$ and $\beta$. However the quartic vertices 
(computed before electroweak symmetry breaking) 
written in terms of the non-physical fields $\varphi_i^\pm$ , 
$h_i$ and $z_i$ (i=1,2) are considerably simpler expressions. 
The crucial point of \cite{kanemura} is the fact that 
the $S$ matrix expressed in terms of the physical fields 
(i.e. the mass eigenstate fields) can be transformed into 
an $S$ matrix for the non-physical fields $\varphi_i^\pm$ , $h_i$ and $z_i$
by making a unitarity transformation. 
The latter is relatively easy to compute from  eq. \ref{higgspot}. 
Therefore the full set of scalar scattering processes can be expressed as 
an $S$ matrix composed of 4 submatrices which 
do not couple with each other due to charge conservation and 
CP-invariance. The entries are the quartic couplings which mediate
the scattering processes.

The first submatrix corresponds to scatterings whose
initial and final states are one of the following:
$(\varphi_1^+\varphi_2^-$,$\varphi_2^+\varphi_1^-$, $h_1 z_2$, $h_2z_1 $, 
 $z_1 z_2$, $h_1h_2)$. 
Therefore one obtains a $6\times 6$ matrix leading to the following 
5 distinct eigenvalues:
\begin{eqnarray}
& & e_1=2 \lambda_3 - \lambda_4 - \frac{\lambda_5}{2} + \frac{5}{2} \lambda_6 \nonumber\\
& & e_2=2 \lambda_3 + \lambda_4 - \frac{\lambda_5}{2} + \frac{1}{2}\lambda_6 \nonumber\\
& & f_+=2 \lambda_3 - \lambda_4 +  \frac{5}{2}\lambda_5 - \frac{1}{2} \lambda_6\nonumber\\
& & f_-=2 \lambda_3 + \lambda_4 + \frac{1}{2}\lambda_5 - \frac{1}{2}\lambda_6\nonumber\\
& & f_1=f_2=2 \lambda_3 + \frac{1}{2} \lambda_5 + \frac{1}{2} \lambda_6
\end{eqnarray}

The second submatrix corresponds to scatterings with initial and final
states one of the following:  
$(\varphi_1^+\varphi_1^-$, $\varphi_2^+\varphi_2^-$, 
$\frac{z_1 z_1}{\sqrt{2}}$, 
 $\frac{z_2z_2}{\sqrt{2}}$, $\frac{h_1 h_1}{\sqrt{2}}$,
$\frac{h_2h_2}{\sqrt{2}})$. Again we obtain a $6\times 6$ matrix with 
the following 6 eigenvalues:
\begin{eqnarray}
& & a_{\pm} =3 (\lambda_1 + \lambda_2 + 2 \lambda_3) \pm
\sqrt{9 (\lambda_1 - \lambda_2)^2 +
      (4 \lambda_3 +  \lambda_4 + \frac{1}{2}(\lambda_5 + \lambda_6))^2}\\
& & b_{\pm}=\lambda_1 + \lambda_2 + 
2 \lambda_3 \pm \sqrt{ (\lambda_1 - \lambda_2)^2 +
\frac{1}{4}(-2 \lambda_4 + \lambda_5 + \lambda_6)^2}\\
& & c_{\pm}=\lambda_1 + \lambda_2 + 
2 \lambda_3 \pm \sqrt{(\lambda_1 - 
\lambda_2)^2 + \frac{1}{4}(\lambda_5 - \lambda_6)^2} 
\end{eqnarray}
The third submatrix corresponds to the basis:
$(h_1 z_1 , h_2 z_2)$. The $2 \times 2$ matrix
possesses the eigenvalues $d_{\pm}$ and $c_{\pm}$, with
$d_{\pm}=c_{\pm}$.
All the above eigenvalues agree with those found in \cite{kanemura},
up to a factor of $1/16\pi$ which we have factorised out.
In our analysis we also include the two body scattering 
between the 8 charged
states: $h_1 \varphi_1^+ $,  $h_2 \varphi_1^+ $,  $z_1 \varphi_1^+ $ , 
$z_2 \varphi_1^+$ ,
$h_1\varphi_2^+$ , $h_2 \varphi_2^+$ , $z_1 \varphi_2^+$, $z_2 \varphi_2^+$.
Note that these channels were neglected in \cite{kanemura}. 
The 8$\times$8 submatrix obtained from the above scattering processes 
contains many vanishing elements, and the 8 eigenvalues are straightforward
to obtain analytically. They read as follows:
 $f_-$, $e_2$ , $f_{1}$, $c_{\pm}$, $b_{\pm}$ and $p_1$,
where
\begin{eqnarray}
p_1 =2 (\lambda_3 +  \lambda_4 ) - \frac{1}{2}\lambda_5 - \frac{1}{2}\lambda_6
\end{eqnarray}
As one can see, these additional channels lead only to one extra
eigenvalue, $p_1$, although we shall see that this eigenvalue plays
an important role in constraining $M_{H^\pm}$ and $M_A$. 

\renewcommand{\theequation}{3.\arabic{equation}}
\setcounter{equation}{0}
\section*{3. Numerical results and discussion}
In this section we present our results for the unitarity constraints
on the Higgs masses in the THDM. All the eigenvalues are
constrained as follows:
\begin{eqnarray}
|a_{\pm}| \ , \ |b_{\pm}| \ , \ |c_{\pm}| \ , \ |d_{\pm}| \ , \ |f_{\pm}| \ , 
\ |e_{1,2}| \ , \ |f_{1,2}| \ , \ |p_1| \ \leq 8\pi
\label{constraint}
\end{eqnarray}
In \cite{kanemura} the eigenvalues explicitly contain the factor $1/16\pi$
and so were constrained to be less than $1/2$.  In section 3.1 we consider 
the special case 
of $\lambda_5=0$, while section 3.2 presents results for the general case
of  $\lambda_5\ne 0$.

\subsection*{3.1 Case of $\lambda_5=0$}
For $\lambda_5=0$ our potential is identical to those considered
in \cite{malampi,kanemura}. We improve those analyses on two accounts:

\begin{itemize}
\item [{(i)}] We have considered extra scattering channels which
leads to one more eigenvalue constraint, $p_1$, as explained in Section 2.

\item [{(ii)}] When finding the allowed parameter space of Higgs masses
we simultaneously impose all the eigenvalue constraints. 
In \cite{kanemura} only the condition $|a_+|\le 8\pi$ was applied when 
deriving mass bounds. 

\end{itemize}
In order to obtain the upper bounds on the Higgs masses
allowed by the unitarity constraints we vary all the Higgs 
masses and mixing angles randomly over a very large parameter space. 
We confirm the result of \cite{kanemura} which states that $a_+$ is 
comfortably the strongest individual eigenvalue constraint. However,
the other eigenvalues impose important constraints on $M_A$ and 
$M_{H^\pm}$. If only $|a_+|\le 8\pi$  
is imposed we can reproduce the upper bounds on the Higgs masses given
in \cite{kanemura}, in particular their main result of $M_h\le 410$ GeV.
When all eigenvalue bounds are applied simultaneously we find improved
bounds on the Higgs masses, particularly for $M_A$ and $M_{H^\pm}$.
We note that the new eigenvalue constraint $p_1\le 8\pi$ (eq.2.9)
plays a crucial role in determining the upper bound on $M_{H^\pm}$. 
We write in Table.1 our bounds (AAN) and those of 
\cite{kanemura}, denoted by KKT. 

\begin{table}
\centering
\begin{tabular} {|c|c|c|c|c|} \hline
 & $M_{H^\pm}$ & $M_A$ & $M_h$ & $M_H$  \\ \hline
AAN & 691 & 695 & 435 & 638  \\ \hline
KKT & 860  & 1220 & 410 & 700 \\ \hline
\end{tabular}
\caption{A comparison of our bounds (AAN) and those of KKT \protect
{\cite{kanemura}}. All masses are in GeV and $\lambda_5=0$.}
\end{table}
Note that the bounds given in Table.1 are obtained for 
relatively small $\tan\beta$ (say $\tan\beta\approx 0.5$). For 
large $\tan\beta$ the bound is stronger, although for the case of
$A^0$, $H^0$ and $H^{\pm}$ the $\tan\beta$ dependence is
rather gentle. Of particular interest is 
the $\tan\beta$ dependence of the bound on $M_h$ which will covered 
in the section 3.2.
\begin{figure}[hbt]
\centerline{\protect\hbox{\psfig{file=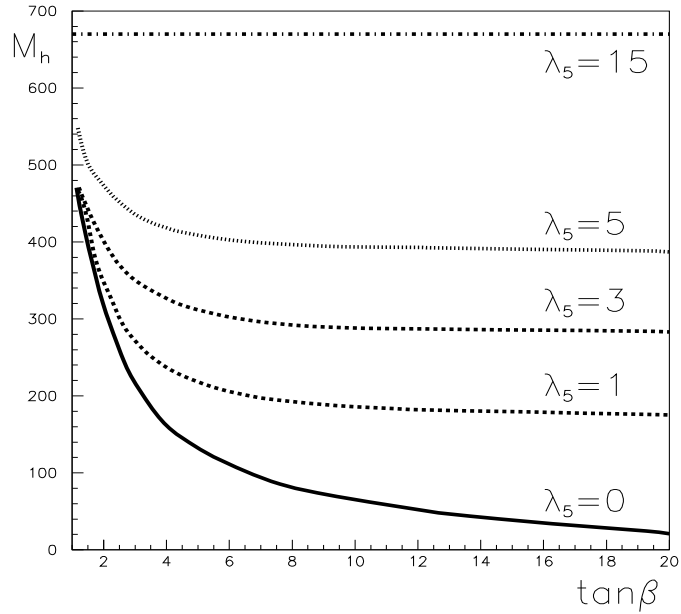,height=10cm,width=10cm}}}
\caption{Maximum $M_h$ in GeV as a function of $\tan\beta$ for various
values of $\lambda_5$.}
\end{figure}
\begin{figure}[htb]
\centerline{\protect\hbox{\psfig{file=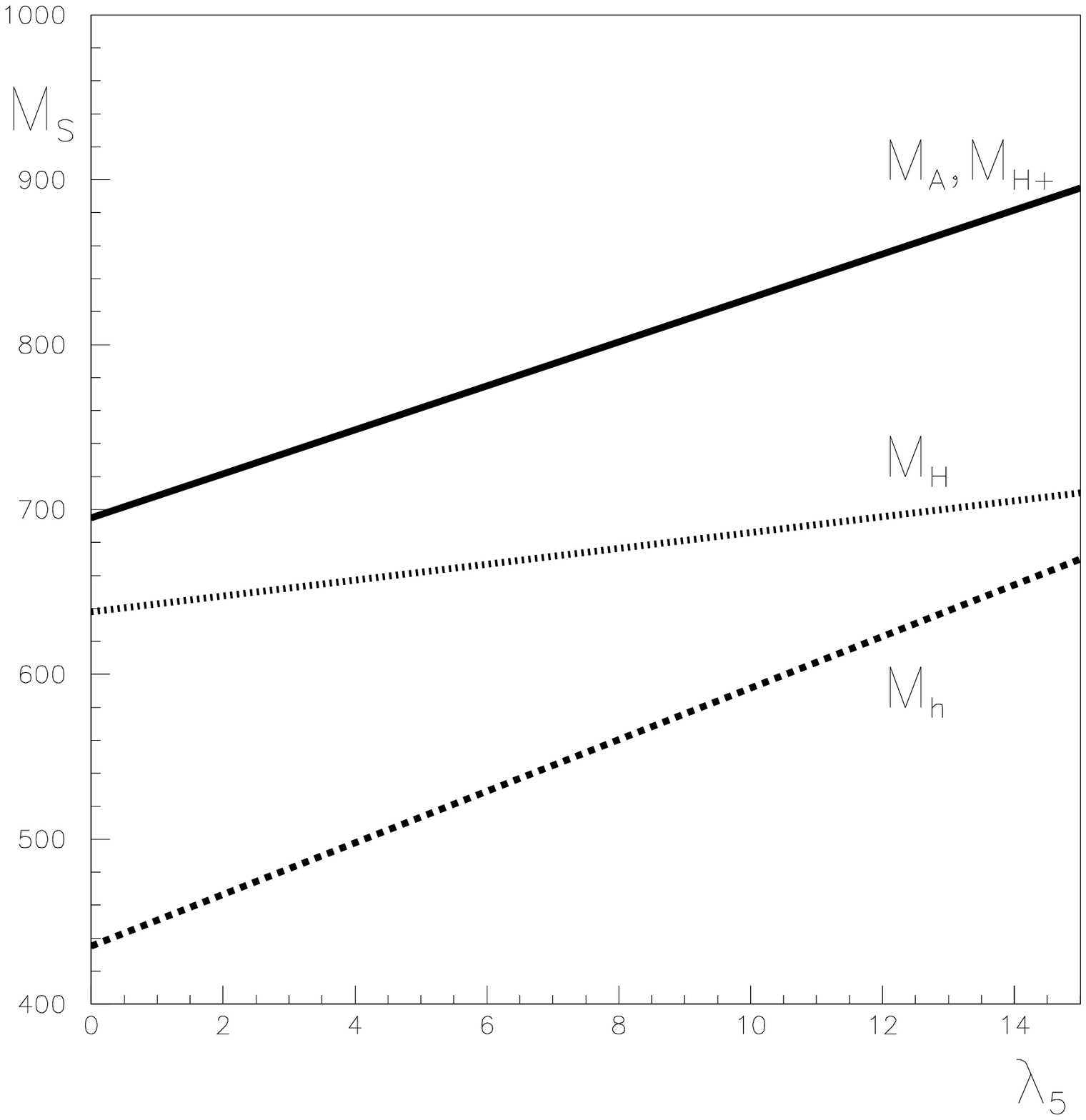,height=10cm,width=10cm}}}
\caption{Maximum mass value of the Higgs scalars ($M_S$) in GeV
as a function of $\lambda_5$.}
\end{figure}
\subsection*{3.2 General case of $\lambda_5\ne 0$}
We now consider $\lambda_5\ne 0$ which corresponds to the 
inclusion of the term which softly breaks the discrete symmetry.
Such a term was neglected in the analyses of \cite{malampi, kanemura},
and from perturbative constraints may take values $|\lambda_5|\le 8\pi$ 
\cite{okada}. In the graphs which follow we do not impose the
perturbative requirement $|\lambda_i|\le 8\pi$ for the remaining 
$\lambda_i$.  Imposing this condition only leads to minor changes in the
numerical results which will be commented on when necessary.

We plot in Fig.1 the maximum value of
$M_h$ against $\tan\beta$ for increasing values of $\lambda_5$, imposing
all the eigenvalue constraints simultaneously as done in Section 3.1. 
For the case of $\lambda_5=0$ one finds a strong correlation, with 
larger $\tan\beta$ requiring smaller $M_h$.
For example, $\tan\beta\ge 7$ corresponds to $M_h\le 100$ GeV, which is 
the mass range already being probed by LEPII.
However, $h^0$ in the THDM with $M_h\le 100$ GeV is not guaranteed to 
be found at LEPII due to the suppression factor of $\sin^2(\beta-\alpha)$
for the main production process $e^+e^-\to h^0Z$.
For the case of $\lambda_5=0$ we find that values of 
$\tan\beta\ge 20$ are strongly disfavoured
since they easily violate one of the unitarity constraints. 
If $\lambda_5\ne 0$, Fig.1 shows that for a given $\tan\beta$,
the action of increasing $\lambda_5$ allows larger maximum values of $M_h$. 
For $\lambda_5=15$ one finds a horizontal line at $M_h\approx 670$ GeV, 
showing that the upper bound has been increased for all values of 
$\tan\beta$. In addition, large ($\ge 30$) values of $\tan\beta$ 
are allowed if $\lambda_5\ne 0$, in contrast to the case
of $\lambda_5=0$. For example, $\lambda_5=1$ 
comfortably permits values of $\tan\beta=60$.
However, as pointed out in \cite{HW} perturbative constraints on the 
$\lambda_i$ also restrict the 
allowed values of $\tan\beta$ in the THDM. Using the condition in
\cite{okada} which requires $|\lambda_i|\le 8\pi$, we found that 
$\tan\beta\ge 30$ is strongly disfavoured. 
In Fig.2 we plot the maximum mass values ($M_S$) of all the Higgs 
bosons as a function
of $\lambda_5$. Again all the Higgs masses and mixing angles have
been varied randomly. We see that $\lambda_5=0$ corresponds to the values
in Table 1, while for $\lambda_5=15$ the upper bounds have been increased
significantly. Including the perturbative requirement would lower the
bounds on $M_A$ and $M_{H^\pm}$ by $10\to 20$ GeV. The $\tan\beta$
dependence of $M_S$ for $S=H^{\pm},A^0$ and $H^0$
is not very pronounced, and the maximum mass value may be obtained for 
both small and large $\tan\beta$.  

We note that the relaxation of the strong correlation between $M_h$ and 
$\tan\beta$ with $\lambda_5\ne 0$ would in principle allow the possibility
of distinguishing between the discrete symmetry conserving and violating 
potentials. If $h^0$ is discovered and the measured values of $M_h$ 
and $\tan\beta$ lie outside the rather constrained region for 
$\lambda_5=0$, this would signify $\lambda_5\ne 0$ and thus a soft 
breaking of the discrete symmetry.
Possibilities of measuring $\tan\beta$ at high--energy $e^+e^-$ colliders
have been considered in the context of the MSSM in \cite{Feng}.
Production and decay of $H^\pm$ and $A^0$ are particularly promising
since the rates do not involve the mixing angle $\alpha$.
Much of the analysis of \cite{Feng} is also valid in the THDM.

\section*{4. Conclusions}
We have derived upper limits on the masses of the Higgs bosons
in the general Two Higgs Doublet Model (THDM) by requiring that unitarity
is not violated in scalar scattering processes. We first considered
the THDM scalar potential which is invariant under a discrete symmetry
transformation and improved previous studies by including the complete
set of scattering channels. Stronger constraints on the Higgs masses
were derived. Of particular interest is the $\tan\beta$ dependence
of the upper bound on $M_h$, with larger $\tan\beta$ requiring
a lighter $h^0$ e.g. $\tan\beta\ge 7$ implies 
$M_h\le 100$ GeV. We then showed that the presence of the discrete symmetry
breaking term parametrized by $\lambda_5$ may significantly weaken the
upper bounds on the masses. In particular, the aforementioned correlation 
between $\tan\beta$ and maximum $M_h$ is relaxed. It was suggested that 
a measurement of $\tan\beta$ and $M_h$ may allow discrimination between
the two potentials.  

\section*{Acknowledgements}
A.G.A was supported by the Japan Society for Promotion of Science (JSPS).
We wish to thank Y. Okada for useful discussions, and C. Dove for reading the
manuscript.

\end{document}